\documentclass[floats, prd, eqnum, showpacs, nofootinbib, 
twocolumn, 
eqsecnum]{revtex4-1}

\usepackage{color,graphicx}
\usepackage{amsfonts}
\usepackage{amssymb}
\usepackage{comment}
\usepackage{ulem}
\begin{document}

\title{Relation between circular photon orbits and the stability of wormholes with the thin shell of a barotropic fluid}
\author{Naoki Tsukamoto${}^{1}$}\email{tsukamoto@rikkyo.ac.jp}
\author{Takafumi Kokubu${}^{2}$}\email{takafumi.kokubu@ipmu.jp}

\affiliation{
${}^{1}$Department of Physics, Faculty of Science, Tokyo University of Science, 1-3, Kagurazaka, Shinjuku-ku, Tokyo 162-8601, Japan \\
${}^{2}$Kavli Institute for the Physics and Mathematics of the Universe (WPI), University of Tokyo, Kashiwa 277-8583, Japan \\
}

\begin{abstract}
We cut a general, static, spherically symmetric spacetime
and paste its copy to make a wormhole with a thin shell of any barotropic fluid in general relativity.
We show that the stability of the thin-shell wormhole is characterized by a set of circular photon orbits called an (anti)photon sphere in the original spacetime
if a momentum flux passing through a throat is prohibited.
Our result will be useful to classify the stability of the thin shell on the throat against linearized spherically symmetric perturbations.
\end{abstract}

\maketitle

\section{Introduction}
Recently, the LIGO and VIRGO Collaborations have detected gravitational waves from compact objects~\cite{Abbott:2016blz}, 
and the Event Horizon Telescope Collaboration has detected the shadows of black hole candidates 
at the centers of a giant elliptical galaxy M87~\cite{Akiyama:2019cqa} and of the Milky Way~\cite{EventHorizonTelescope:2022wkp}.
The study of compact objects with a strong gravitational field in theoretical and observational aspects will be important to understanding our Universe.

Static, spherically symmetric compact objects such as black holes and wormholes have unstable (stable) circular photon orbits called photon (antiphoton) spheres~\cite{Claudel:2000yi,Perlick_2004_Living_Rev}.
Photon spheres have important roles in several phenomena in a strong gravitational field.
Dim images near compact objects~\cite{Hagihara_1931,Darwin_1959,Atkinson_1965}, 
the image of a collapsing star to become a black hole~\cite{Ames_1968,Synge:1966okc},
the photon absorption cross section~\cite{Sanchez:1977si}, 
quasinormal modes~\cite{Press:1971wr,Goebel_1972},
centrifugal force and gyroscopic precession~\cite{Abramowicz_Prasanna_1990,Abramowicz:1990cb,Allen:1990ci,Hasse_Perlick_2002},
and
Bondi's sonic horizon of a radial fluid~\cite{Mach:2013gia,Chaverra:2015bya,Cvetic:2016bxi,Koga:2016jjq,Koga:2018ybs,Koga:2019teu} are related to the photon sphere.
Recently, the features of a circular photon orbit like  
its radius~\cite{Hod:2017xkz}, its number~\cite{Cunha:2017eoe,Hod:2017zpi}, and its stability~\cite{Cunha:2017eoe,Koga:2019uqd} 
have been investigated.

A wormhole is a hypothetical spacetime structure with nontrivial topology which is permitted in general relativity~\cite{Visser_1995,Morris:1988cz}.
The wormhole connects two regions of one universe or two universes by its throat.
The wormhole solutions should be stable in order for wormholes to exist in nature.
In Refs.~\cite{Visser:1989kg,Poisson:1995sv}, the Schwarzschild spacetime is cut and, 
its two copies are pasted by a thin shell~\cite{Lanczos:1922,Israel:1966rt} to construct a wormhole solution by Darmois-Israel matching~\cite{Darmois:1927,Israel:1966rt},
and the stability of this thin-shell wormhole is studied.
After Refs.~\cite{Visser:1989kg,Poisson:1995sv}, the stability of thin-shell wormholes pasted by using
general static, spherically symmetric spacetimes~\cite{Eiroa:2008ky,Garcia:2011aa,Lobo:2005zu,Ishak:2001az},
plane symmetric spacetimes~\cite{Lemos:2008aj},
asymmetric spacetimes~\cite{Garcia:2011aa,Eid:2017xcg,Tsukamoto:2021fpp},
cylindrical spacetimes~\cite{Eiroa:2004at},
higher-dimensional spacetimes~\cite{Thibeault:2005ha}, 
and lower-dimensional spacetimes~\cite{Perry:1991qq}
were investigated.
We note that the stability of wormholes depends on gravitational theories~\cite{Rahaman:2007bf}.
Recently, the details of the stability of a traversable thin-shell wormhole have been investigated in Refs.~\cite{Nakao:2013hba,Akai:2017aro}. 

In 2000, Barcelo and Visser~\cite{Barcelo:2000ta} cut a class of static, spherically symmetric spacetime 
at a radius and pasted two copies of the spacetime to make a thin-shell wormhole.
They pointed out that the location of the static throat filled with a pure tension $\sigma=-p$, 
where $\sigma$ and $p$ are the surface energy density and the surface pressure of the thin shell, 
was equal to the radius of photon sphere associated with the original spacetime.
Recently, Koga~\cite{Koga:2020gqd} showed that the throat of a general static pure-tensional thin-shell wormhole with only $Z_2$ symmetry in $\Lambda$-vacuum is located on a photon surface~\cite{Claudel:2000yi}
and showed that the stability of the thin shell corresponds to the stability of the photon surface.
These studies imply that (anti)photon spheres in the original spacetime, before being cut and pasted, may have a relation in the stability of a thin-shell wormhole beyond the pure tension case,
and this relation can be useful to classify the stability of the thin-shell wormholes.

In this paper, we consider a general, static, $Z_2$-symmetrical, and spherically symmetric wormhole
with a thin shell filled with any barotropic fluid $p=p(\sigma)$.
We find that the stability of the thin-shell wormhole under linearized spherically symmetric perturbations has a relation with (anti)photon spheres in general relativity 
if a momentum flux passing through the throat is forbidden.

We consider the general spherical metric to study the stability of the thin-shell wormhole
but we should keep in mind that the stability analysis of the thin-shell wormhole will be valid 
only if the original spacetime satisfies a generalized Birkhoff's theorem~\cite{Eiroa:2008ky,Bronnikov1980,Schleich:2009uj}. 
In cases where the theorem cannot be applied, the spherically symmetric perturbations to the thin shell 
can affect metrics outside of the thin shell to emit gravitational waves. 

This paper is organized as follows:
In Sec.~II, we review the photon sphere and the antiphoton sphere.
In Sec.~III, we make a general, static, spherically symmetric wormhole with a thin shell filled with any barotropic fluid,
and we show a relation between (anti)photon spheres and the (in)stability of the thin-shell wormhole if the momentum flux passing through the throat is prohibited.
We give examples in Sec.~IV and we discuss and conclude our results in Sec.~V.
In this paper, we use units in which light speed and Newton's constant are unity.

\section{Photon sphere and antiphoton sphere}
We consider a general, static, spherically symmetric spacetime with a line element 
\begin{eqnarray}
ds^2=-A(r)dt^2+B(r)dr^2+ C(r) \left( d\theta^2 +\sin^2 \theta d\phi^2  \right), \nonumber\\
\end{eqnarray}
where $A(r)$, $B(r)$, and $C(r)$ are functions of a radial coordinate $r$.
We assume that $A(r)$, $B(r)$, 
$C(r)$, and $C^{\prime}(r)$ 
are positive and finite in a range $r \geq a$, where $a$ is a constant and $'$ is a differentiation with respect to the radial coordinate $r$.
There are time-translational and axial Killing vectors $t^\mu \partial_\mu=\partial_t$ and $\phi^\mu \partial_\mu=\partial_\phi$ 
because of the stationarity and axisymmetry of the spacetime, respectively.
From spherical symmetry, we can assume $\theta=\pi/2$ without loss of generality.

The motion of a light ray is described by
\begin{eqnarray}
-A(r)dt^2+B(r)dr^2+ C(r) d\phi^2 =0
\end{eqnarray}
and it is rewritten in
\begin{eqnarray}
\left(\frac{dr}{d\lambda}\right)^2 +\mathcal{V}(r) =0,
\end{eqnarray}
where $\lambda$ is an affine parameter on the trajectory of the ray 
and $\mathcal{V}(r)$ is an effective potential of the motion of the ray defined by
\begin{eqnarray}
\mathcal{V}(r) \equiv \frac{1}{B} \left( \frac{L^2}{C}-\frac{E^2}{A} \right),
\end{eqnarray}
where 
\begin{eqnarray}
E\equiv -g_{\mu \nu} t^\mu \frac{dx^\nu}{d\lambda}=A \frac{dt}{d\lambda}
\end{eqnarray}
 and 
\begin{eqnarray}
L\equiv g_{\mu \nu} \phi^\mu \frac{dx^\nu}{d\lambda}=C \frac{d\phi}{d\lambda}
\end{eqnarray}
are the conserved energy and angular momentum of the light ray, respectively.
The first and second derivatives of $\mathcal{V}$ with respect to the radial coordinate $r$ are given by
\begin{equation}
\mathcal{V}'=
-\frac{B'}{B}\mathcal{V}
-\frac{1}{B}\left( \frac{C'}{C}L^2 - \frac{A'}{A}E^2 \right)
\end{equation}
and 
\begin{eqnarray}
\mathcal{V}'' 
&=&-\left( \frac{B'}{B} \right)' \mathcal{V}- \frac{B'}{B}\mathcal{V}'
+\frac{B'}{B^2}\left( \frac{C'}{C^2}L^2- \frac{A'}{A^2}E^2 \right) \nonumber\\
&&+\frac{2}{B}\left( \frac{C'^2}{C^3}L^2- \frac{A'^2}{A^3}E^2 \right)
-\frac{B'}{B^2}\left( \frac{C''}{C^2}L^2- \frac{A''}{A^2}E^2 \right), \nonumber\\
\end{eqnarray}
respectively.

The circular orbit of the ray with a radius $r=r_\mathrm{m}$ should satisfy a condition 
\begin{equation}
\mathcal{V}_\mathrm{m}=\mathcal{V}'_\mathrm{m}=0
\end{equation}
and it gives 
\begin{equation}
L^2=L^2_\mathrm{m}\equiv \frac{C_\mathrm{m}}{A_\mathrm{m}}E^2
\end{equation}
and 
\begin{equation}
D_\mathrm{m}=0,
\end{equation}
where 
\begin{equation}\label{eq:D}
D(r)\equiv \frac{A'}{A}-\frac{C'}{C}.
\end{equation}
Here and hereafter, a function with a subscript $m$ denotes the function on the circular orbit at $r=r_\mathrm{m}$.
We obtain 
\begin{eqnarray}\label{eq:ddV}
\mathcal{V}''_\mathrm{m} =\frac{E^2}{A_\mathrm{m}B_\mathrm{m}} F_\mathrm{m},
\end{eqnarray}
where 
\begin{eqnarray}\label{eq:F}
F(r) \equiv \frac{A''}{A}- \frac{C''}{C}, 
\end{eqnarray}
and the circular orbit of the light is stable if $F_\mathrm{m}>0$, and it is unstable if $F_\mathrm{m}<0$.
The unstable (stable) circular orbit is called the photon sphere (antiphoton sphere).

\section{Thin-Shell wormhole}
We construct a thin-shell wormhole by cutting and pasting a general, static, spherically symmetric spacetime \cite{Darmois:1927,Israel:1966rt}. 
We assume that the original spacetime satisfies a generalized Birkhoff's theorem~\cite{Eiroa:2008ky,Bronnikov1980,Schleich:2009uj} and that spherically symmetric perturbations to a thin shell 
do not affect metrics outside of the thin shell. 
We take two copies of a manifold $\Omega_\pm \equiv \left\{ r_{\pm} > a   \right\}$
with boundaries given by timelike hypersurfaces $\Sigma_\pm \equiv \left\{ r_{\pm} = a  \right\}$.
We identify the hypersurfaces $\Sigma \equiv \Sigma_+ = \Sigma_- $ 
to obtain a manifold $\mathcal{M}$ by gluing the manifolds~$\Omega_\pm$ at a throat located at $\Sigma$.
Note that the wormhole has $Z_2$ symmetry against the throat.
The hypersurface~$\Sigma$ filled with Dirac distribution matter is called a thin shell.
Coordinates in $\Omega_\pm$ denote $x^\mu$,  
but the coordinates may not joint continuously at the two-dimensional hypersurface~$\Sigma$. 
We denote by $y^i$ coordinates on the two-dimensional hypersurface~$\Sigma$.
We assume that the same coordinates $y^i$ can be taken on both sides of the hypersurface~$\Sigma$.
We permit $a=a(\tau)$, where $\tau$ is its proper time, since we are interested in the dynamics of the thin shell. 

We consider that the hypersurface $\Sigma$ is orthogonally sticked by a congruence of geodesics. The geodesics are
parametrized by the proper distance $l$, and we set $l=0$
when the geodesics intersect the hypersurface and $l<0$ $(l>0)$
when they are in $\Omega_-$ $(\Omega_+)$.
A displacement from the hypersurface $\Sigma$ is given by $dx^{\mu} =n^{\mu}dl$, where $n^{\mu}$ is the unit normal to the hypersurface.
A metric tensor in $\mathcal{M}$ is given by
$g_{\mu \nu} = \Theta(-l) g_{- \mu \nu}+ \Theta (l) g_{+\mu \nu}$, where $\Theta(l)$
is the Heaviside distribution, which is  $0$ if
$l<0$, and 1 if $l>0$, and which is indeterminate if
$l=0$, and where $g_{-\mu \nu}$ and $g_{+\mu \nu}$ are metric tensors in $\Omega_-$ and $\Omega_+$, respectively.

The connection $\Gamma^\mu_{\nu \rho}$ is given by
\begin{eqnarray}
\Gamma^\mu_{\nu \rho}=\Theta(-l) \Gamma^\mu_{-\nu \rho} + \Theta(l) \Gamma^\mu_{+\nu \rho}, 
\end{eqnarray}
where $\Gamma^\mu_{-\nu \rho}$ and $\Gamma^\mu_{+\nu \rho}$ are the connections in $\Omega_-$ and $\Omega_+$, respectively.
The extrinsic curvature $K_{ij}$ of the timelike hypersurface $\Sigma$ is given by
\begin{equation}\label{eq:K_ij}
K_{ij}
\equiv  e^\mu_i e^\nu_j \nabla_\nu  n_\mu
=(n_{\mu,\nu}-\Gamma^\rho_{\mu \nu} n_{\rho}) e^\mu_i e^\nu_j,
\end{equation}
where $e^\mu_i$ are basis vectors:
\begin{eqnarray}
e^\mu_i \equiv 
\frac{\partial  x^{\mu}}{\partial y^{i}},
\end{eqnarray}
and $\nabla_\nu$ is a covariant differentiation in $\mathcal{M}$.

The induced metric $h_{ij}\equiv g_{\mu \nu} e^{\mu}_{i} e^{\nu}_{j}$ on the hypersurface $\Sigma$ can be written by
\begin{eqnarray}
ds_{\Sigma}^2
&=&h_{ij}dy^idy^j \nonumber\\
&=&-d\tau^2+C(a) \left( d\theta^2 +\sin^2 \theta d\phi^2  \right).
\end{eqnarray}
The induced metrics in $\Omega_-$ and $\Omega_+$ are the same as each other.
The four-velocity $u^\mu$ of the thin shell at $t=T(\tau)$ and $r=a(\tau)$ is given by 
\begin{eqnarray}
u^\mu \partial_\mu=\dot{T} \partial_t+\dot{a} \partial_r,
\end{eqnarray}
where the dot denotes a differentiation with respective to $\tau$ and 
\begin{eqnarray}
\dot{T} = \sqrt{\frac{1+B\dot{a}^2}{A}}.
\end{eqnarray}
We obtain
\begin{eqnarray}
\ddot{T}= \frac{2\ddot{a}\dot{a}AB+\dot{a}^3(AB'-A'B)-\dot{a}A'}{2A^2\dot{T}}.
\end{eqnarray}

The unit normals $n_{\mu \pm}$ to the hypersurface in $\Omega_-$ and $\Omega_+$ are obtained as
\begin{equation}
n_{\mu \pm} dx^\mu= \pm \left( -\sqrt{AB}\dot{a} dt +\sqrt{AB}\dot{T} dr \right)
\end{equation}
and the basis vectors $e^\mu_i$ are given by
\begin{eqnarray}
e^\mu_\tau \partial_\mu &=&\dot{T} \partial_t + \dot{a} \partial_r, \\
e^\mu_\theta \partial_\mu &=& \partial_\theta, \\
e^\mu_\phi \partial_\mu &=& \partial_\phi.
\end{eqnarray}
The extrinsic curvatures of the hypersurfaces in $\Omega_\pm$ are given by 
\begin{eqnarray}\label{eq:K_tautau}
K^\tau_{\tau \pm}&=& \frac{\pm 1}{\sqrt{\dot{a}^2+\frac{1}{B}}}\left( \ddot{a}+\frac{\dot{a}^2 (AB)'}{2AB }+\frac{A'}{2AB } \right), \\\label{eq:K_thetatheta}
K^\theta_{\theta \pm}&=&K^\phi_{\phi \pm}= \frac{\pm C'}{2C}\sqrt{\dot{a}^2+\frac{1}{B}}, 
\end{eqnarray}
and the traces are  
\begin{equation}\label{eq:K}
K_\pm \equiv \frac{\pm 1}{\sqrt{\dot{a}^2+\frac{1}{B}}}\left( \ddot{a}+\frac{\dot{a}^2 (AB)'}{2AB }+\frac{A'}{2AB } \right) \pm \frac{C'}{C}\sqrt{\dot{a}^2+\frac{1}{B}}.
\end{equation}
The Einstein equations, which the thin shell should satisfy, are given by 
\begin{equation}\label{eq:Einstein_eq}
S^i_j= - \frac{1}{8\pi} \left( \left[ K^i_j \right] - \left[ K \right] \delta^i_j \right),
\end{equation} 
where $S^i_j$ is a surface stress-energy tensor for the thin shell 
\begin{equation}\label{eq:S_ij}
S^i_j=(\sigma+p)U^iU_j+p \delta^i_j,
\end{equation}
where we define $U_i dy^i \equiv u_\mu e^\mu_i dy^i= d\tau$, 
and where $\sigma$ and $p$ are the surface energy density and the surface pressure of the thin shell, respectively
and we obtain $S^\tau_\tau=-\sigma$ and $S^\theta_\theta=S^\phi_\phi=p$.
Here, $\left[ T \right]$ is defined by
\begin{equation}
\left[ T \right] \equiv \left. T_+ \right|_{\Sigma} - \left. T_- \right|_{\Sigma},
\end{equation}
where $T_+$ and $T_-$ are any tensorial function $T$ in $\Omega_+$ and $\Omega_-$, respectively.
From $(\tau,\tau)$ and $(\theta,\theta)$ components of the Einstein equations~(\ref{eq:Einstein_eq}), we obtain the surface energy density $\sigma$ and the surface pressure $p$: 
\begin{equation}\label{eq:S^tau_tau}
\sigma= - \frac{1}{4\pi} \frac{C'}{C}\sqrt{\dot{a}^2+\frac{1}{B}}
\end{equation}
and
\begin{equation}\label{eq:S^phi_phi}
p=\frac{1}{8\pi}\frac{1}{\sqrt{\dot{a}^2+\frac{1}{B}}} \left( 2\ddot{a} + \frac{\dot{a}^2 \left( AB C \right)'+\left( AC \right)'}{AB C} \right),
\end{equation}
respectively.
From Eqs.~(\ref{eq:S^tau_tau}) and (\ref{eq:S^phi_phi}), we obtain
\begin{equation}\label{eq:Energy_conservation0}
\frac{d( \sigma \mathcal{A})}{d\tau} +p\frac{d \mathcal{A}}{d\tau}=-\frac{\dot{a}}{2} \sqrt{\dot{a}^2+\frac{1}{B(a)}} C'(a) H(a),
\end{equation}
where $\mathcal{A} \equiv 4 \pi C(a)$ is the area of the throat. 
Here, we have defined $H(r)$ by 
\begin{equation}
H(r) \equiv \frac{2C^{\prime \prime}(r)}{C'(r)}-\frac{\left(A(r)B(r) C(r) \right)' }{A(r)B(r) C(r)}. 
\end{equation}

A term on the right-hand side of Eq.~(\ref{eq:Energy_conservation0}) means a momentum flux passing through the throat. 
Under the assumptions that $\dot{a}$ is nonzero, $C^{\prime}(a)$ is positive, and $B(a)$ has a positive and finite value, 
the flux term vanishes if and only if $H(a)=0$ holds.
In this paper, we name $H(r)=0$, the no-flux-term condition, which was also investigated in Refs.~\cite{Eiroa:2008ky,Jacobson:2007tj}.
Note that we have defined that the no-flux-term condition as a condition of the original spacetime before being cut and pasted.
If we transform the radial coordinate $r$ to be $A(r)B(r)=1$ in the original spacetime without loss of generality,
we can integrate the no-flux-term condition $H(r)=0$ and we obtain 
\begin{eqnarray}\label{eq:C1}
C(r)=(c_1 r+c_2)^2,
\end{eqnarray}
where $c_1$ and $c_2$ are integral constants. Equation~(\ref{eq:C1}) means that the radial coordinate is an affine parameter. For simplicity, we assume that $c_1$ is positive and $c_2$ is not negative.   
If the original spacetime satisfies  
\begin{eqnarray}\label{eq:noheatflow}
A(r)B(r)=1,\quad   C(r)=(c_1 r+c_2)^2,
\end{eqnarray}
then the no-flux-term condition $H(r)=0$ holds.

Equation~(\ref{eq:Energy_conservation0}) is rewritten in
\begin{equation}\label{eq:Energy_conservation2}
C(a)\sigma'+C'(a)(\sigma +p) =\frac{C(a)}{2} H(a) \sigma,
\end{equation}
where $\sigma' \equiv \dot{\sigma}/\dot{a}$.
If we assume a barotropic fluid with $p=p(\sigma)$, from Eq.~(\ref{eq:Energy_conservation2}), we obtain the surface density $\sigma=\sigma(a)$.
From Eq.~(\ref{eq:S^tau_tau}), the equation of motion for the thin shell is given by
\begin{equation}
\dot{a}^2+V(a)=0,
\end{equation}
where $V(a)$ is an effective potential defined by
\begin{equation}
V(a)\equiv \frac{1}{B}-\left( \frac{4\pi \sigma C}{C'} \right)^2.
\end{equation}
The derivative of $V$ with respect to $r$ is given by
\begin{equation}
V'= -\frac{B'}{B^2}-\frac{32\pi^2 \sigma C}{C'} \left( \left( 1-\frac{CC''}{C'^2} \right) \sigma + \frac{C \sigma'}{C'} \right),
\end{equation}
and from Eq.~(\ref{eq:Energy_conservation2}), it can be rewritten as
\begin{equation}
V'= -\frac{B'}{B^2}+\frac{16\pi^2 \sigma C}{C'} \left( 2p + \frac{(AB C)'}{AB C'}\sigma \right).
\end{equation}
The second derivative of $V$ is obtained as
\begin{eqnarray}
&&V'' \nonumber\\
&&= -\frac{B''B-2B'^2}{B^3} \nonumber\\
&&+16\pi^2 \left\{ \left( \frac{C}{C'} \sigma' + \left( 1- \frac{CC''}{C'^2} \right) \sigma \right) \left( 2p +\frac{(AB C)'}{AB C'} \sigma \right) \right. \nonumber\\
&& + \frac{C}{C'} \sigma \left( 2p' + \frac{(AB C)'}{AB C'} \sigma' \right. \nonumber\\
&& \left. \left. +\frac{((AB)'' AB - (AB)'^2)C'- AB (AB)' C''}{(AB C')^2}C \sigma \right) \right\}, \nonumber\\
\end{eqnarray}
and from Eq.~(\ref{eq:Energy_conservation2}), it becomes
\begin{eqnarray}
&&V''=\nonumber\\
&& -\frac{B''B-2B'^2}{B^3} 
-8\pi^2 \left\{ \left( 2p +\frac{(AB C)'}{AB C'} \sigma \right)^2 \right. \nonumber\\
&&+\sigma \left( 2p+ \left( 2+ \frac{(AB C)'}{AB C'}- \frac{2CC''}{C'^2} \right) \sigma \right) \left( \frac{(AB C)'}{AB C'}+2\beta^2 \right) \nonumber\\
&&\left. +\frac{2C^2 \left( \left( (AB)'^2-AB (AB)'' \right) C' +AB (AB)' C'' \right) \sigma^2}{(AB)^2 C'^3} \right\}, \nonumber\\
\end{eqnarray}
where $\beta^2\equiv dp/d\sigma=p'/\sigma'$.

We consider a static thin shell at $a=a_0$, where $a_0$ is a positive constant, with the surface energy density 
\begin{equation}
\sigma_0= - \frac{C_0'}{4\pi\sqrt{B_0}C_0} 
\end{equation}
and the surface pressure 
\begin{equation}
p_0=\frac{\left( A_0C_0 \right)'}{8\pi\sqrt{B_0} A_0C_0},
\end{equation}
where the subscript~$0$ denotes the functions at $a=a_0$.
From the definition, $V_0=V'_0=0$ is satisfied. 
Therefore, the effective potential can be expanded around $a=a_0$ as
\begin{equation}
V(a)=\frac{V_0''}{2}(a-a_0)^2 +O\left( \left( a-a_0 \right)^3 \right), 
\end{equation}
where $V_0''$ is given by
\begin{eqnarray}\label{eq:ddV_0}
V_0''
&=&\frac{1}{B_0} \left( 2G_0 \beta_0^2 +\frac{A_0''}{A_0} +G_0  \right. \nonumber\\
&& \left. -\frac{A_0'^2}{A_0^2}-\frac{A_0'B_0'}{2A_0B_0}-\frac{A_0'C_0'}{A_0C_0}-\frac{B_0'C_0'}{B_0C_0}\right),
\end{eqnarray}
where we define $G(r)$ as
\begin{eqnarray}\label{eq:G}
G(r)
&\equiv& \frac{C''}{C} -\frac{B'C'}{2BC}-\frac{C'^2}{C^2} \nonumber\\
&=&\frac{C^\prime}{2 C} \left( D +H \right).
\end{eqnarray}
When $V_0''>0$ $(V_0''<0)$, the thin shell is stable (unstable).

If we assume the no-flux-term condition $H(r)=0$ and we choose the radial coordinate to be $A(r)B(r)=1$ in the original spacetime, 
then we obtain
\begin{eqnarray}\label{eq:ddV0}
V_0''
&=&A_0 \left( \frac{2}{c_1a_0+c_2}D_0 \beta^2_0  +F_0 -\frac{A_0'}{2A_0}D_0 \right).
\end{eqnarray}
The thin shell is stable if 
\begin{eqnarray}
\beta^2_0 > \frac{c_1a_0+c_2}{2 D_0} \left( -F_0 +\frac{A_0'}{2A_0}D_0 \right)
\end{eqnarray}
for $D_0>0$,
and if
\begin{eqnarray}
\beta^2_0 < \frac{c_1a_0+c_2}{2 D_0} \left( -F_0 +\frac{A_0'}{2A_0}D_0 \right)
\end{eqnarray}
for $D_0<0$.
Note that an antiphoton sphere or a photon sphere is found at $D_{\mathrm{m}}=0$. 
Thus, we realize that the existence of the antiphoton sphere and photon sphere of the original spacetime 
strongly affects the stability of the thin-shell wormhole spacetime. 
If $D_{\mathrm{m}0}=0$ holds, 
we obtain
\begin{eqnarray}\label{eq:ddVm0}
V_{\mathrm{m}0}''
=A_{\mathrm{m}0}F_{\mathrm{m}0}.
\end{eqnarray}
Here, the subscript~``m$0$'' denotes the functions at $r=a_0=r_\mathrm{m}$.
Therefore, the thin shell on the antiphoton spheres (photon spheres) is stable (unstable) independent of the value of $\beta^2_{\mathrm{m}0}$.

\section{Application}
We apply our formulas to the Reissner-Nordstr\"{o}m spacetime and the Kottler spacetime, which is often called Schwarzschild-(anti-)de Sitter spacetime.
We review not only the (anti)photon spheres, but also subextremal and extremal event horizons and cosmological horizons in the original 
Reissner-Nordstr\"{o}m and Kottler spacetimes.
This is because we cannot make static wormholes 
by using the regions of the inside of the event horizons and the outside of the cosmological horizon in the original spacetimes,
and because a strong gravitational field, which can almost form the cosmological horizons and the subextremal event horizons, would make the throat unstable.
Thus, we can categorize the stability of the thin-shell wormholes by considering not only the existence of the photon and antiphoton spheres, 
but also the existence of the event and cosmological horizons in the original spacetimes.
        
\subsection{Reissner-Nordstr\"{o}m spacetime}
The Reissner-Nordstr\"{o}m spacetime is a static, spherically symmetric, asymptotically flat electrovacuum solution in general relativity with  
\begin{eqnarray}
A&=&\frac{1}{B}=1-\frac{2M}{r}+\frac{Q^2}{r^2}\\
C&=&r^2
\end{eqnarray}
and it satisfies the condition~(\ref{eq:noheatflow}).
The stability of the wormhole produced by cutting the Reissner-Nordstr\"{o}m spacetimes and sticking them together with the thin shell of the barotropic fluid  
was investigated by Eiroa and Romero in 2004~\cite{Eiroa:2003wp}.

\begin{figure*}[htbp]
\begin{center}
\includegraphics[width=85mm]{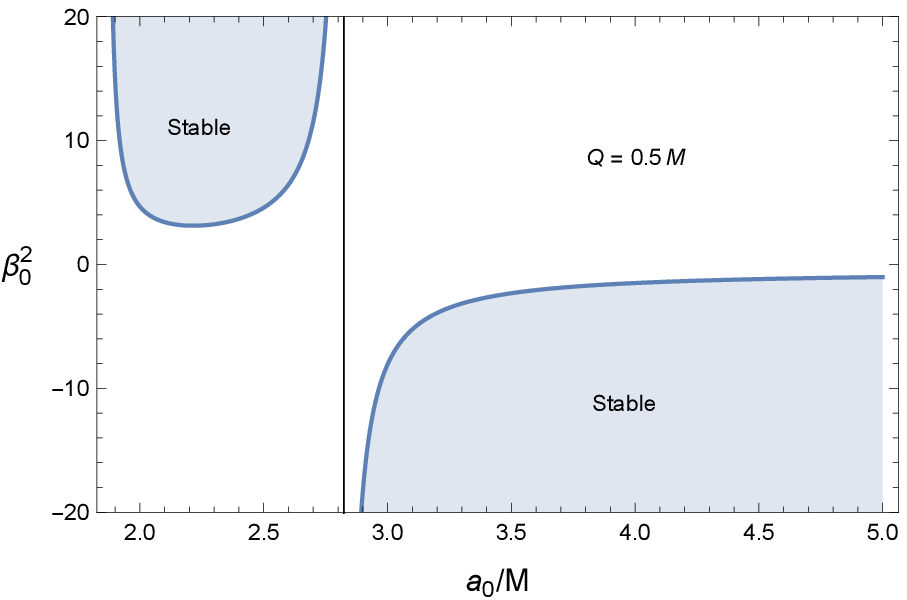}
\includegraphics[width=85mm]{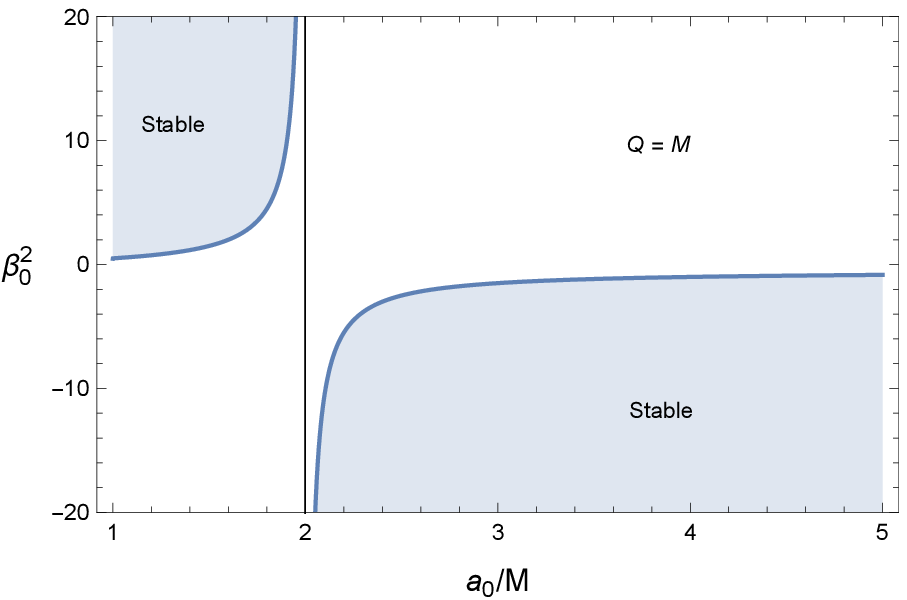}
\includegraphics[width=85mm]{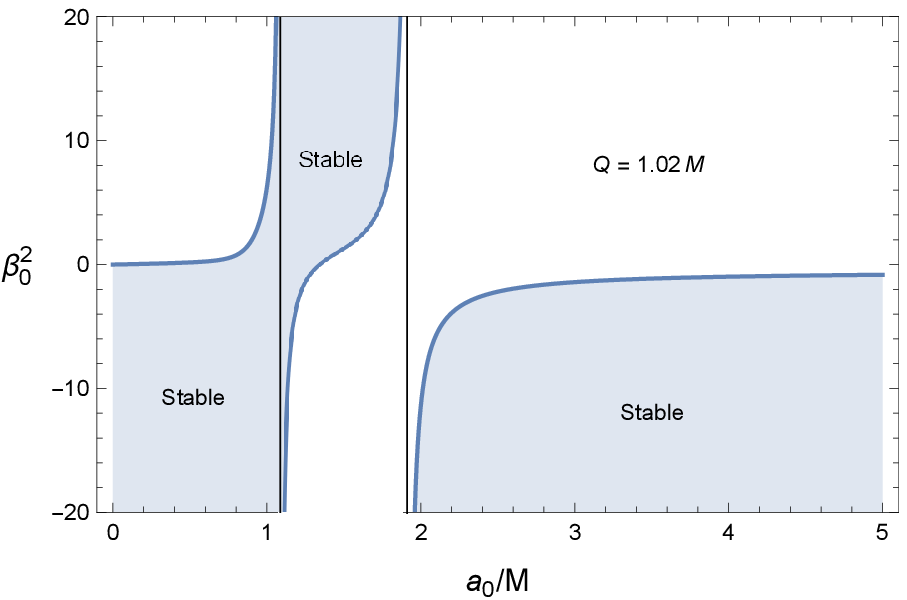}
\includegraphics[width=85mm]{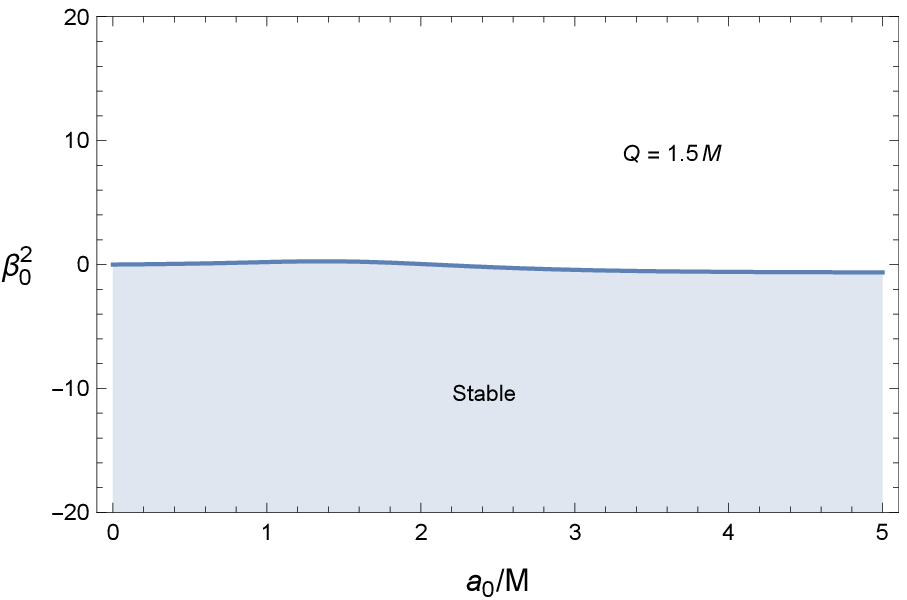}
\end{center}
    \caption{Stability of the Reissner-Nordstr\"{o}m wormhole for $Q=0.5M$ (top-left, case~I), $Q=M$ (top-right, case~II), $Q=1.02M$ (bottom-left, case~III), and $Q=1.5M$ (bottom-right, case~V). Colored parts are stable.
    For $Q=0.5M$ (top-left, case~I), the original Reissner-Nordstr\"{o}m spacetime has a photon sphere at $r=2.82M$ and a subextremal event horizon at $r=1.87M$.  
    For $Q=M$ (top-right, case~II), it has a photon sphere at $r=2M$ and an extremal event horizon at $r=M$.  
    For $Q=1.02M$ (bottom-left, case~III), it has a photon sphere at $r=1.91M$ and an antiphoton sphere at $r=1.09M$ and no event horizon.
    For $Q=1.5M$ (bottom-right, case~V), it has no the photon sphere, antiphoton sphere, or event horizon.
    }\label{RN}
\end{figure*}

We can categorize the spacetime into case~I ($\left| Q \right| < M$), where it has a subextremal event horizon 
at $r=M +\sqrt{M^2 - Q^2}$ and a photon sphere at $r=(3M+\sqrt{9M^2-8Q^2})/2$;
case~II ($\left| Q \right| = M$), where it has an extremal event horizon at $r=M$ and a photon sphere at $r=2M$; 
case~III ($M < \left| Q \right| < 3\sqrt{2}M/4$), where it has the photon sphere and an antiphoton sphere at $r=(3M-\sqrt{9M^2-8Q^2})/2$ and no event horizon;
case~IV ($\left| Q \right| = 3\sqrt{2}M/4$), where it has a marginal unstable photon sphere~\cite{Tsukamoto:2020iez} and no event horizon; 
and case~V ($\left| Q \right| > 3\sqrt{2}M/4$), where it has no the photon sphere, antiphoton sphere, or event horizon. 
The stability in cases~I-III and V are shown in Fig.~\ref{RN}. We do not show the stability in the case~IV, since it falls outside of our assumptions. 

\subsection{Kottler spacetime}
\begin{figure*}[htbp]
\begin{center}
\includegraphics[width=85mm]{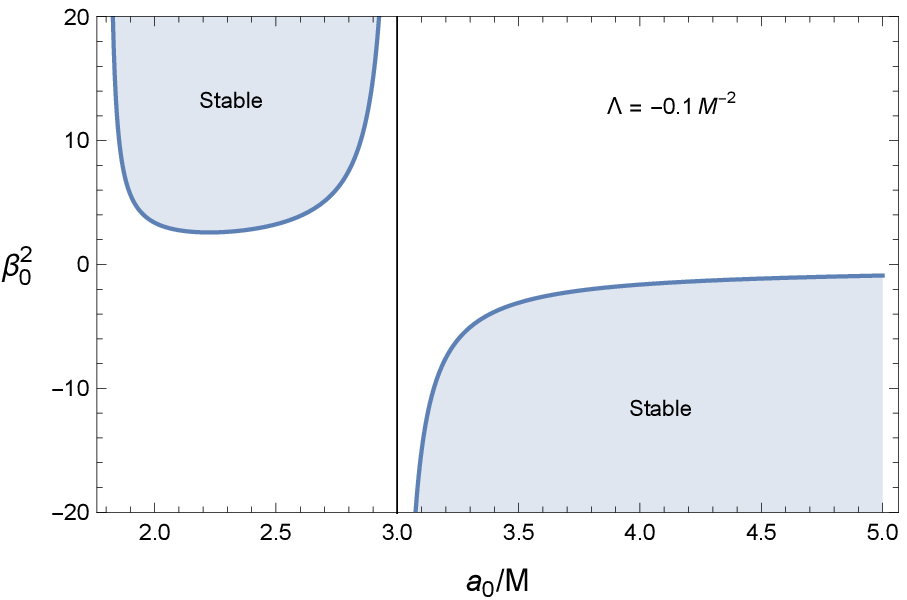}
\includegraphics[width=85mm]{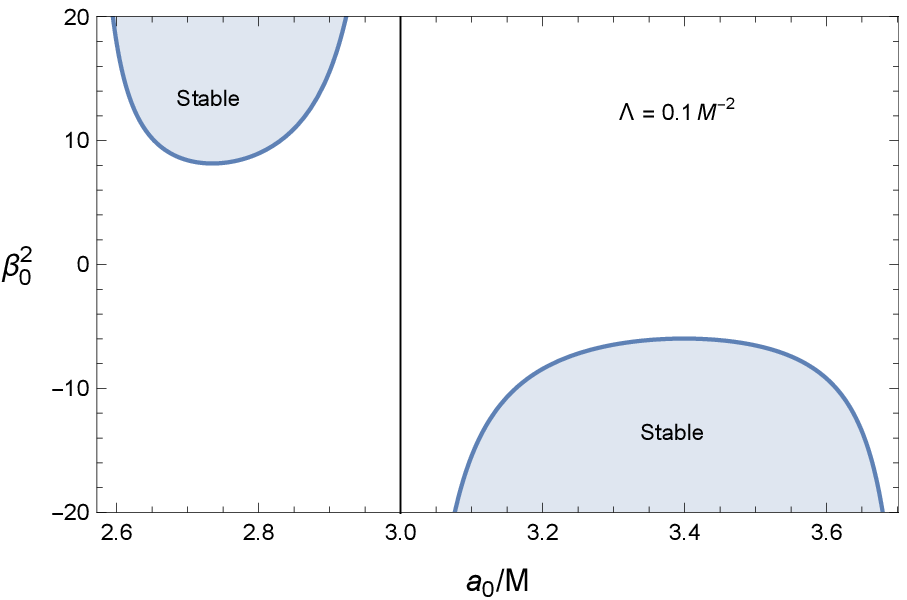}
\end{center}
    \caption{Stability of the Kottler wormhole for $\Lambda=-0.1M^{-2}$ (left, case~I) and $\Lambda=0.1M^{-2}$ (right, case~VI). Colored parts are stable.
    For $\Lambda=-0.1M^{-2}$ (left, case~I), it has a photon sphere at at $r=3M$ and a subextremal event horizon at $r=r_\mathrm{e}=1.80M$.
    For $\Lambda=0.1M^{-2}$ (right, case~VI), the original Kottler spacetime has a photon sphere at $r=3M$, the event horizon at $r=r_\mathrm{e}=2.56M$, and a cosmological horizon at $r=r_\mathrm{c}=3.73M$. 
    }\label{K}
\end{figure*}

The Kottler [Schwarzschild-(anti-)de Sitter] spacetime is the unique spherically symmetric, vacuum solution of Einstein equations with a cosmological constant $\Lambda$. Its metric has   
\begin{eqnarray}
A&=&\frac{1}{B}=1-\frac{2M}{r}-\frac{\Lambda r^2}{3}\\
C&=&r^2
\end{eqnarray}
and it satisfies the condition~(\ref{eq:noheatflow}).
The stability of the thin-shell Kottler wormhole with the thin shell of the barotropic fluid  
was investigated by Lobo and Crawford in 2004~\cite{Lobo:2003xd}. 
We can categorize it into case~I ($\Lambda \leq 0$), where it has a subextremal event horizon at $r=r_\mathrm{e}$ and a photon sphere at $r=3M$,
and case~VI [$0<\Lambda<(3M)^{-2}$], where it has a subextremal event horizon, a photon sphere, and a cosmological horizon at $r=r_\mathrm{c}$. 
It has a static region only in $r_\mathrm{e}<r<r_\mathrm{c}$.
Note that $r_\mathrm{e}$ and $r_\mathrm{c}$ satisfy the relation $2M<r_\mathrm{e}<3M<r_\mathrm{c}$.
For $(3M)^{-2}\leq \Lambda$, it has no static region. See Ref.~\cite{Podolsky:1999ts}
for the details of an extreme case, $\Lambda=(3M)^{-2}$. 
Figure~\ref{K} shows the stability in the cases~I and VI.

\section{Discussion and Conclusion}
Recently, interesting features of (anti)photon spheres and circular photon orbits not only in wormhole spacetimes, but also in various other spacetimes have been investigated.
Shaikh \textit{et al.}~\cite{Shaikh:2019itn} pointed out that the images of ultracompact objects with a photon sphere and an antiphoton sphere,
such as the interior Schwarzschild solutions due to Synge~\cite{Synge} and Florides~\cite{Florides}, can be brighter than 
the images of objects with a photon sphere and no antiphoton sphere. 
Kudo and Asada prove that a spacetime cannot be asymptotically flat if its outermost circular photon orbit is stable~\cite{Kudo:2022ewn}.
The relation between the (in)stability of spacetimes and circular photon orbits has been also investigated eagerly.  
Stable and unstable circular photon orbits might cause the instability of ultracompact neutron stars, boson stars, Proca stars, and so on~\cite{Keir:2014oka,Cardoso:2014sna,Cunha:2017qtt,Cunha:2022gde},
while we may find linearly stable ultracompact objects if they do not have ergoregions~\cite{Zhong:2022jke}.

In this paper, we have cut a general, static, spherically symmetric spacetime
and joined two copies to make a thin-shell wormhole spacetime satisfying the condition 
of a thin shell filled with any barotropic fluid in general relativity. We have shown that the stability and instability of the throat strongly 
depend on the existence of the photon sphere and antiphoton sphere of the original spacetime
under the assumption that the original spacetime satisfies a generalized Birkhoff's theorem~\cite{Eiroa:2008ky,Bronnikov1980,Schleich:2009uj},
and that a momentum flux passing through the throat is forbidden.

Stable wormholes without the thin shell in general relativity have not been found to date~\cite{Shinkai:2002gv,Gonzalez:2008wd,Gonzalez:2008xk,Gonzalez:2009hn},
but Bronnikov~\textit{et al.} have reported a candidate of a stable wormhole with some matter sources~\cite{Bronnikov:2013coa}, 
and Azad~\textit{et al.} have discussed that the rotation of a wormhole may stabilize it~\cite{Azad:2023iju}.
We notice that there are few studies on the (in)stability analysis of a wormhole with an antiphoton sphere on a throat,
while the instability of static and spherically symmetrical wormholes with a photon sphere on the throat have been
reported often.

Moreover, in a general context, the possibility that static, spherically symmetrical, and $Z_2$-symmetrical wormholes can have an antiphoton sphere on a throat has been overlooked often.   
For example, Bronnikov and Baleevskikh have shown that a general static, spherically symmetrical, and $Z_2$-symmetrical wormhole has circular photon orbits on the throat from the symmetry, 
and they have concluded that the wormhole has a photon sphere on the throat~\cite{Bronnikov:2018nub}. 
On the other hand, in Ref.~\cite{Shaikh:2019jfr}, Shaikh~\textit{et al.} have pointed out that wormholes can have an antiphoton sphere on the throat,
and Tsukamoto has shown that a Damour-Solodukhin wormhole~\cite{Damour:2007ap} or a Bronnikov-Kim wormhole~\cite{Bronnikov:2002rn,Bronnikov:2003gx} can have an antiphoton sphere on the throat in Refs.~\cite{Tsukamoto:2020uay,Tsukamoto:2021apr}.

In this paper, we have concentrated on the thin-shell wormhole with $Z_2$ symmetry against the throat,
but our result gives us a method to find stable wormholes without the thin shell: Our result implies 
that the antiphoton sphere on or near the throat might stabilize wormholes. 
Thus, the investigation of wormholes with an antiphoton sphere on or near their throats 
could be a good strategy to find stable wormholes without the thin shell.

\section*{Acknowledgements}
We are deeply grateful to an anonymous referee who provided valuable comments.
This work was partially supported by JSPS KAKENHI Grants No. JP20H05853 (T.K.) from the Japan Society for the Promotion of Science.
%

\end{document}